\documentclass[referee]{raa}

\usepackage{graphicx,times}            
\usepackage{natbib}
\usepackage{amssymb,amsmath}
\usepackage{longtable}
\bibpunct{(}{)}{;}{a}{}{,}

\usepackage[pagebackref=true]{hyperref}
\usepackage{caption}
\begin{document}
	
	\title{The year-scale X-ray variations in the core of M87
	}

	\volnopage{Vol.0 (20xx) No.0, 000--000}      
	\setcounter{page}{1}          
	
	\author{Yu-Lin Cheng 
		\inst{1}, Fei Xiang\inst{1}\thanks{This paper is dedicated to the memory of my wonderful tutor Fei Xiang, who recently passed away.}
		, Heng Yu\inst{2}, Shu-Mei Jia\inst{3}, Xiang-Hua Li\inst{1}, Cheng-Kui Li\inst{3}, Yong Chen\inst{3}, Wen-Cheng Feng\inst{4}
	}

	\institute{Department of Astronomy, Yunnan University
		Kunming 650091, China; {\it chengyl@mail.ynu.edu.cn}; {\it xhli@ynu.edu.cn}\\
		\and
		Department of Astronomy, Beijing Normal University, Beijing, 100875, China \\
		\and
		Key Laboratory of Particle Astrophysics, Institute of High Energy Physics, Chinese Academy of Sciences, Beijing 100049, China\\
		\and
		Department of Physics and Institute of Theoretical Physics, Nanjing Normal University, Nanjing 210023, China\\
		\vs\no
		{\small Received 20xx month day; accepted 20xx month day}}
	
	\abstract{ The analysis of light variation of M87 can help us understand the disc evolution.
		In the past decade, M87 has experienced several short-term light variabilities related to flares. 
		We also find there are year-scale X-ray variations in the core of M87. Their light variability properties are similar to clumpy-ADAF. 
		By re-analyzing 56 {\it Chandra} observations from 2007 to 2019, we distinguish the `non-flaring state' from `flaring state' in the light variability.
		After removing flaring state data, we identify 4 gas clumps in the nucleus and all of them can be well fitted by the clumpy-ADAF model. The average mass accretion rate is $\sim 0.16 \rm M_{\odot}  yr^{-1}$.
		We analyze the photon index($\Gamma$) — flux(2-10keV) correlation between the non-flaring state and flaring state.  
		For the non-flaring states, the flux is inversely proportional to the photon index. For the  flaring states, we find no obvious correlation between the two parameters. 
		In addition, we find that the flare always occurs at a high mass accretion rate, and after the luminosity of the flare reaches the peak, it will be accompanied by a sudden decrease in luminosity. Our results can be explained as that the energy released by magnetic reconnection destroys the structure of the accretion disc, thus the luminosity decreases rapidly and returns to normal levels thereafter.
	}
	
	\authorrunning{Y.-L. Cheng et al. }            
	\titlerunning{The year-scale X-ray variations of M87}  
	
	\maketitle

	\section{Introduction}           
	\label{sect:intro}
	
	M87(NGC4486) is a large radio galaxy located in the Virgo Cluster 
	\citep{Macchetto+etal+1997}
	at a distance of 18.5 Mpc from us 
	\citep{Blakeslee+etal+2001}.
	Its central “engine” is a super-massive black hole (SMBH) with a mass of 
	about $6.5 \times 10^9 \rm M_\odot$ 
	\citep{EHT+2019}.
	M87 emits a high-energy plasma jet extending about 5000 
	light-years from the core, and its relativistic jet is misaligned by an angle of $\sim30^{\circ}$ 
	with respect to our line of sight 
	\citep{Bicknell+etal+1996}.
	The jet components of M87 can be resolved in radio, optical/UV and X-ray bands.
	Considering its large inclination, it is an ideal case
	to study the accretion disc of black hole and the details near the jet.
	
	The radiation mechanism of M87 has been discussed by many researchers. 
	\citet{Wilson+etal+2002} assumed that the X-ray radiation came from the standard thin disc. 
	With a canonical radiation efficiency $\sim0.1$ \citep{Matteo+etal+2000}, the predicted nuclear luminosity of M87 should be $\sim5 \times 10^{44} {\rm erg\ s^{-1}}$ \citep{Matteo+etal+2003}.
	However, the luminosity observed by {\it Chandra} is $\sim7 \times 10^{40} {\rm erg\ s^{-1}}$ \citep{Matteo+etal+2003}. It means that the actual radiation efficiency is $\sim10^{-5}$, four orders magnitude lower than the canonical value,  
	and the required value of radiation efficiency was consistent with the prediction of the Advection Dominated Accretion Flow (ADAF, \citealt{Narayan+etal+1995}) models . 
	\citet{Matteo+etal+2003} fitted the X-ray spectra of M87 with ADAF models and verified that its X-ray radiation was dominated by ADAF. 
	
	As a Low Luminosity AGNs (LLAGNs, \citealt{Yuan+Narayan+2014}), M87 does not have strong flux like blazars. 
	However, from the past observations, it was found that M87 has experienced several short-term light variabilities. 
	In 2005, H.E.S.S captured a TeV emission of M87 with timescales of a few days (\citealt{Aharonian+etal+2006}). 
	The joint observation of {\it Chandra} found a giant flare (\citealt{Harris+etal+2006}) accompanied by this TeV event in knot HST-1 ($0.86^{\prime \prime}$ away from the core). 
	Therefore, \citet{Cheung+etal+2007} proposed HST-1 as a candidate for TeV emission. 
	However, another TeV flare was observed by H.E.S.S., MAGIC (\citealt{Albert+etal+2008}) and VERITAS (\citealt{Acciari+etal+2008}) in 2008 which lasted for about two weeks. 
	In the following days, {\it Chandra} observations suggested that the X-ray intensity of the nucleus was 2 to 3 times higher than usual (\citealt{Harris+etal+2009}). 
	Different to the first outburst, HST-1 was in a low state at this time and its X-ray flux was lower than that in the nucleus. 
	In 2010, a third VHE $\gamma$-ray burst was captured and the time-scale of intensity-doubling was day-scale (\citealt{Aliu+etal+2012}). 
	After the TeV emission, X-ray intensity in the nucleus was also enhanced. Therefore,  it can be confirmed that the site of the TeV flare is the nucleus rather than HST-1 (\citealt{Harris+etal+2011}).
	It is still unclear where the X-ray flare originates. 
	Similar to the solar flare, the X-ray flares of M87 may be triggered by magnetic reconnection (\citealt{Aschwanden+2011}; \citealt{Yuan+etal+2009}; \citealt{Yang+etal+2019}) or from the mini-jet (\citealt{Giannios+etal+2009}). 
	For the intraday variability of the M87 core in 2017, \citet{Imazawa+etal+2021} suggested that the emission might come from the inverse Compton scattering in the jet.

	Previous studies mainly focused on these striking flare events in day-scale or month-scale, 
	but the research on the year-scale light variability of M87 was rare.
	For LLAGNs, the emission of year-scale variation is considered to be related to the accretion mode of the disc. 
	\citet{Wang+etal+2012} proposed that the inhomogeneous accretion flow in LLAGNs might be clumpy (i.e., clumpy-ADAF), 
	which originated from the thermal instability in the accretion flow or is affected by gravity. 
	Once the clump is formed, it will fall toward the centre of the black hole under the tidal force
	and bring about a long-term light variation. 
	By re-analyzing {\it Chandra} observations from 2007 to 2008, \citet{Xiang+Cheng+2020} found a year-scale X-ray variation in the core of M87, and successfully fitted the spectra with a simple clumpy accretion model.
	
	To validate the clumpy-ADAF model, we check the M87 observations of {\it Chandra} from 
	2007 to 2019 and obtain the long-term X-ray variation of M87. 
	We distinguish the `non-flaring state' from `flaring state' in the light curve with a universal classification method.
	Based on the work of \citet{Xiang+Cheng+2020}, we find another three year-scale variability components and reproduce them with a clumpy accretion model.
	This paper is developed as follows: the {\it Chandra} data analysis is described in Section 2, 
	the clumpy accretion model fitting results are presented in Section 3, 
	we discuss the physical characteristics of the clump in Section 4 and finally, conclusions are listed in Section 5.
	The distance $r$ of M87 used in this study is 18.5 Mpc. 
	
	\section{Data and analysis}
	\label{sect:Obs}
	
	To study the long-term X-ray variation of M87, we select the data of M87 observed by {\it Chandra} X-Ray Observatory with subarcsecond resolution. From July 31th, 2007 to March 28th, 2019, 56 observations are carried out using the Advanced CCD Imaging Spectrometer (ACIS) and back-illuminated S3 detector. The time period of each observation is about 4.7ks and the observation mode is FAINT. A 0.4 sec frame time is set to minimize the significant pileup effect (\citealt{Harris+etal+2006}). We use CIAO (version 4.13) to analyse the {\it Chandra} data retrieved from the archive. 
	First of all, we reprocess the data by the \texttt{chandra\_repro} script to ensure that the latest calibrations is consistent with the current version of CIAO.
	
\begin{figure}
		\centering
		\includegraphics[width=10cm, angle=0]{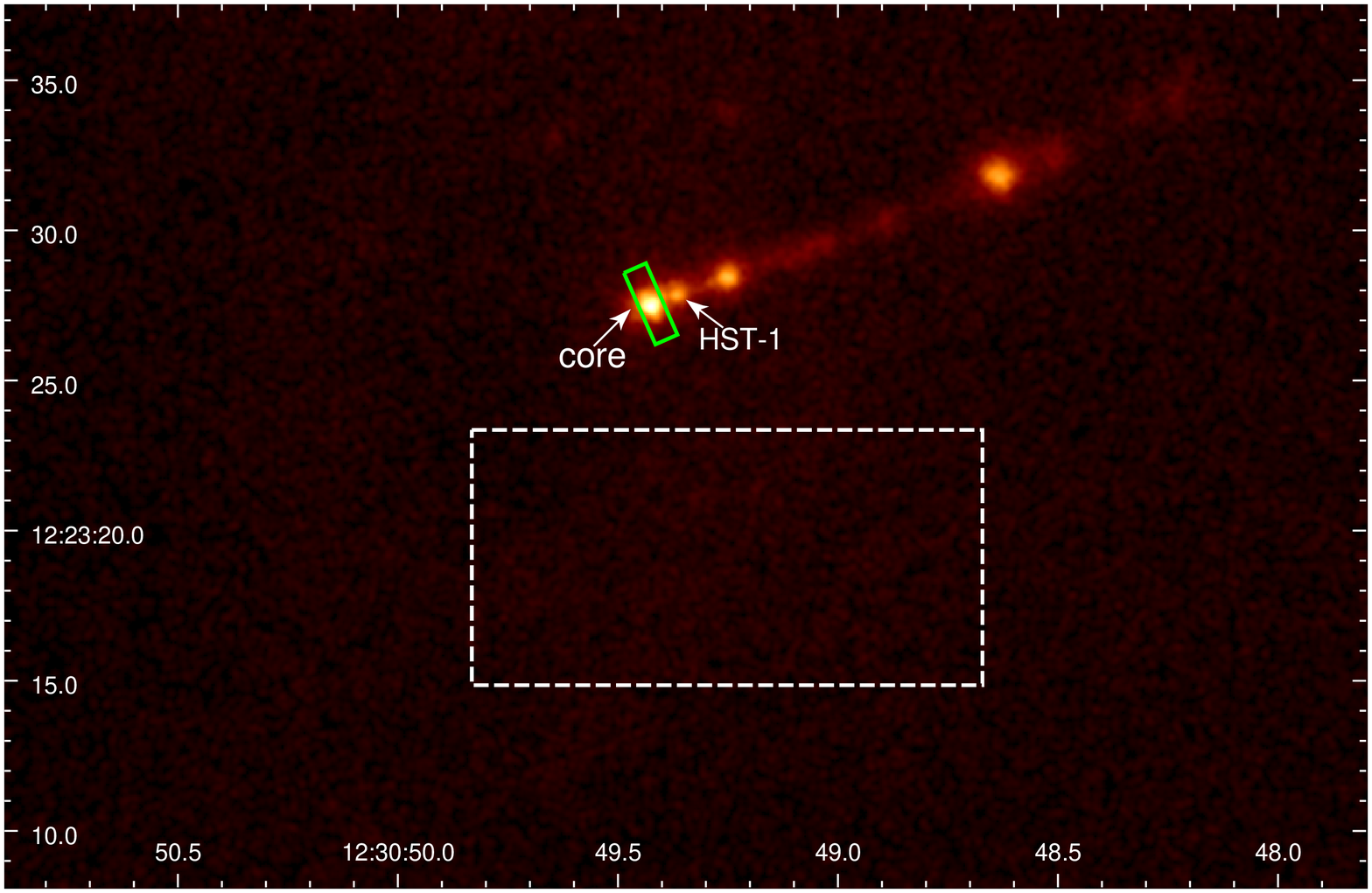}
		\includegraphics[width=11.0cm, angle=0]{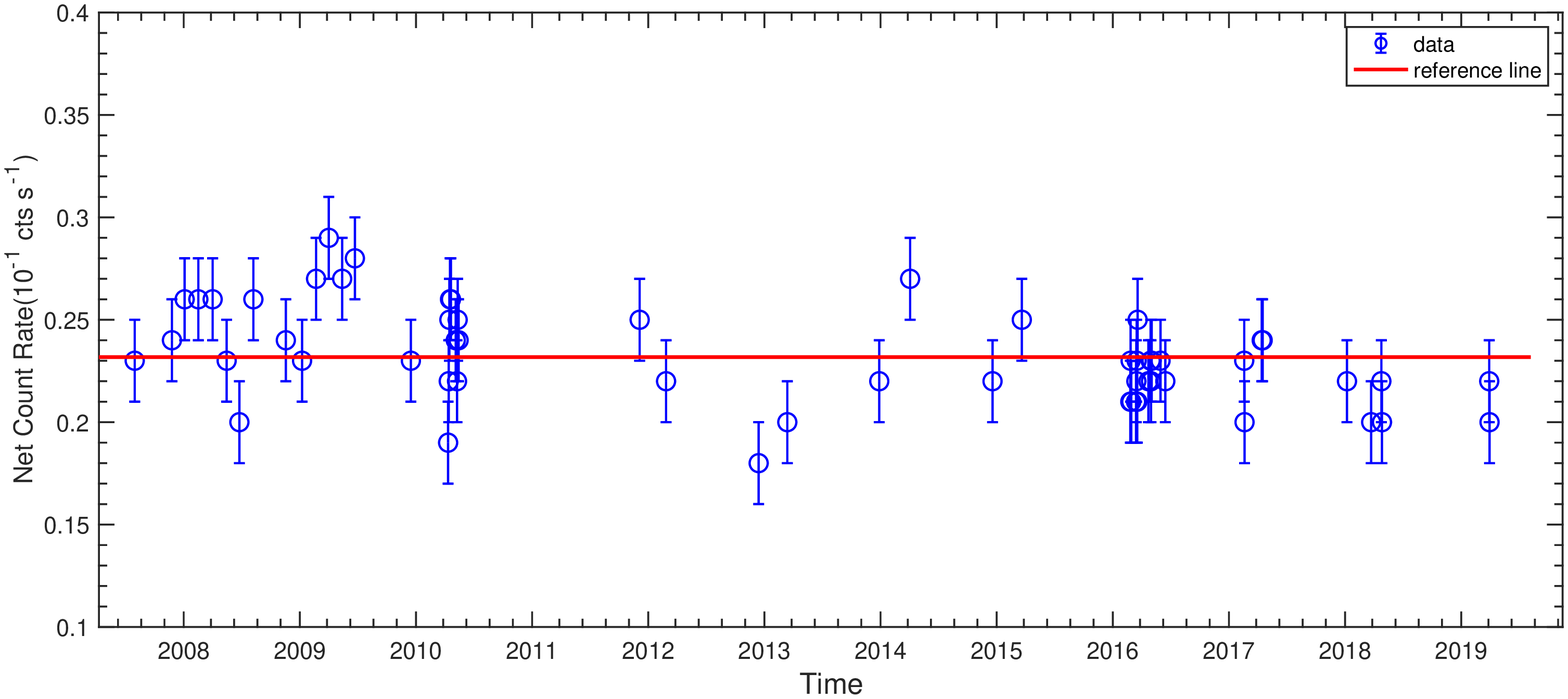}
		\caption{ {\it The top panel}: the X-ray image of M87 observed by {\it Chandra} on February 24th, 2016 (obsID 18781). The images are binned on a scale of 1/8 native ACIS pixel and then smoothed with a Gaussian of FWHM = $0.5^{\prime \prime}$.  The  box in green solid line is the selected source region for the core, and the rectangular in white dotted line is the background region. {\it The bottom panel}: the variation of photon count rate of background region over time of all observations. The solid red line represents the average value.}
		\label{Fig1}
	\end{figure}

	As the core is very close to HST-1, only $0.86^{\prime \prime}$, sometimes the two regions cannot be well distinguished. When HST-1 is bright, possible “light pollution” from HST-1 might happen in the nucleus region, especially for the outburst event of HST-1 in 2005 (\citealt{Harris+etal+2009}; \citealt{Harris+etal+2011}). 
	\cite{Xiang+Cheng+2020}'  analysis showed the nucleus region is little influenced by HST-1 from 2007 to 2008. In 2008, the nucleus was brighter than HST-1, and then the luminosity of HST-1 continued to decrease. Therefore, the “light pollution” of HST-1 on the nucleus can be ignored in our analysis, but we are still careful for border of the core region which might influence the result of our spectra analysis. We adopt a box region with a size of $0.8^{\prime \prime} \times 2.6^{\prime \prime}$ including nucleus (\citealt{Yang+etal+2019}). Since the core is spherical, we take the brightest center of the core as the center of the box region (shown in the top panel of Fig.~\ref{Fig1}).
	
	The surface of ACIS has accumulated a layer of contamination over the mission (\citealt{Plucinsky+etal+2018}).  Since our data spans 12 years, we check the stability of the instrument during this long-term observations. We take a rectangle region with a size of $17^{\prime \prime} \times 8.5^{\prime \prime}$  without resolved point sources as the background region, whose  center is located at  R.A.=$\rm 12^h30^m49.25^s$, Dec=$12^{\circ}23^{\prime}19.10^{\prime\prime}$ (J2000) (shown in the top panel of Fig.~\ref{Fig1}). The photon count rate of the background region of all observations between 2.0-10.0 keV are nearly stable as shown in the bottom panel of Fig.~\ref{Fig1}. 
	Their average value is about 0.023$\pm$0.002 $\rm cts \ s^{-1}$ (the red line in the bottom panel of Fig.~\ref{Fig1}). 
	Thus we omit the instrument contamination at hard X-ray band of 2.0-10.0 keV.

	\section{Results}
	\label{sect:results}
	
	\subsection{Energy Spectra Fitting}
	
\begin{figure}
		\centering
		\includegraphics[width=16cm, angle=0]{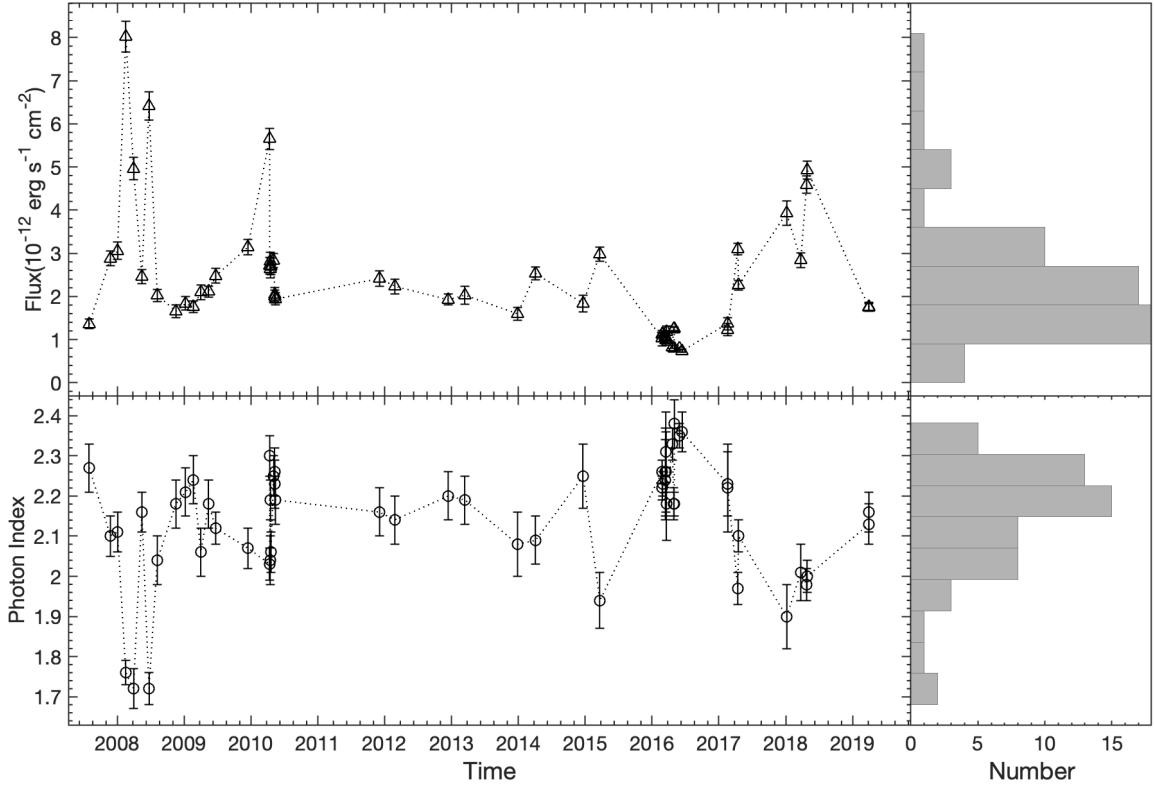}
		\caption{ {\it The top left panel}: the long-term X-ray light curve of M87 from 2007 to 2019.  {\it The bottom left panel}:  the photon index of the core from the spectral fitting result corresponding to the above panel. The upper right panel and the lower rigth panel are the histogram distribution of flux and photon index respectively.}
		\label{Fig2}
	\end{figure}

	The X-ray from jet features of M87 is supposed to come from synchrotron emission and can be characterized by a power law (\citealt{Harris+Krawczynski+2002}; \citealt{Wilson+etal+2002}; \citealt{Harris+etal+2003}). In XSPEC (version 12.1.1), we use a power-law model with Galactic absorption to fit the nuclear X-ray spectra (\citealt{Arnaud+1996}; \citealt{Xiang+Cheng+2020}):
	\begin{equation}\label{eq1}
		{\rm Model=Wabs \ast Powerlaw}.
	\end{equation}
	The column density of hydrogen ($n_{\rm H}$) is fixed as $6.1 \times 10^{20}$ cm$^{-2}$ (\citealt{Wilson+etal+2002}; \citealt{Xiang+Cheng+2020}). We obtained photon index ($\Gamma$), normalization of power-law and flux in 2.0-10.0 keV and the results are listed in Table~\ref{Tab1}. All the reduced chi-squares are less than 1.20 which indicates that our spectra fitting results are reliable. The long-term X-ray light curve of the core is shown in Fig.~\ref{Fig2}.  The variation of the flux intensity  is similar to the light curve in previous works (\citealt{Harris+etal+2009}; \citealt{MAGIC+etal+2020}). 
	
	{
		\small
		\setlength{\tabcolsep}{10pt}
		\begin{longtable}{llllll}
			\caption{List of the spectra fitting results of the core. The flux is in 2.0-10.0 keV and the error of those parameters are calculated with the confidence of 68\%. From July 31th, 2007 to March 28th, 2019, 56 observations are carried out via Chandra/ACIS.}\label{Tab1}    \\     
			\hline\noalign{\smallskip}
			obsID &      Time    & Photon index      & norm              & flux$_{2-10{\rm KeV}}$    &   $\chi^2$/DOF      \\
			&       (MJD)  &                            & (10$^{-5}$)  &  (10$^{-12}$ erg s$^{-1}$ cm$^{-2}$ )  &        \\
			\hline\noalign{\smallskip}
			\endfirsthead
			\multicolumn{6}{c}{Table 1: continued.} \\
			\hline
			obsID &      Time    & Photon index      & norm              & flux$_{2-10{\rm KeV}}$    &   $\chi^2$/DOF      \\
			&       (MJD)  &                            & (10$^{-5}$)  &  (10$^{-12}$ erg s$^{-1}$ cm$^{-2}$ )  &        \\
			\hline\noalign{\smallskip}
			\endhead
			\hline\noalign{\smallskip}
			\endfoot
			\hline\noalign{\smallskip}
			\endlastfoot
			7354	&	54312	&	2.27 	$\pm$	0.06 	&	78	$\pm$	3	&	1.36 	$\pm$	0.11 	&	22.39/39	\\
			8575	&	54429	&	2.10 	$\pm$	0.05 	&	130	$\pm$	4	&	2.88 	$\pm$	0.17 	&	107.06/108	\\
			8576	&	54469	&	2.11 	$\pm$	0.05 	&	14	$\pm$	4	&	3.06 	$\pm$	0.20 	&	27.16/33	\\
			8577	&	54512	&	1.76 	$\pm$	0.03 	&	216	$\pm$	5	&	8.02 	$\pm$	0.36 	&	58.36/56	\\
			8578	&	54557	&	1.72 	$\pm$	0.05 	&	126	$\pm$	4	&	4.96 	$\pm$	0.26 	&	43.42/47	\\
			8579	&	54601	&	2.16 	$\pm$	0.05 	&	121	$\pm$	4	&	2.46 	$\pm$	0.16 	&	122.41/106	\\
			8580	&	54641	&	1.72 	$\pm$	0.04 	&	163	$\pm$	4	&	6.41 	$\pm$	0.33 	&	49.67/46	\\
			8581	&	54685	&	2.04 	$\pm$	0.06 	&	84	$\pm$	3	&	2.02 	$\pm$	0.14 	&	94.01/79	\\
			10282	&	54787	&	2.18 	$\pm$	0.06 	&	85	$\pm$	3	&	1.66 	$\pm$	0.15 	&	53.60/77	\\
			10283	&	54838	&	2.21 	$\pm$	0.06 	&	99	$\pm$	3	&	1.84 	$\pm$	0.16 	&	89.28/88	\\
			10284	&	54882	&	2.24 	$\pm$	0.06 	&	97	$\pm$	3	&	1.76 	$\pm$	0.14 	&	14.14/23	\\
			10285	&	54922	&	2.06 	$\pm$	0.06 	&	89	$\pm$	3	&	2.10 	$\pm$	0.17 	&	18.58/22	\\
			10286	&	54964	&	2.18 	$\pm$	0.06 	&	108	$\pm$	4	&	2.11 	$\pm$	0.13 	&	108.54/91	\\
			10287	&	55004	&	2.12 	$\pm$	0.04 	&	117	$\pm$	4	&	2.48 	$\pm$	0.17 	&	82.84/103	\\
			10288	&	55180	&	2.07 	$\pm$	0.05 	&	143	$\pm$	5	&	3.14 	$\pm$	0.18 	&	38.38/34	\\
			11512	&	55297	&	2.03 	$\pm$	0.04 	&	231	$\pm$	5	&	5.65 	$\pm$	0.25 	&	99.46/105	\\
			11513	&	55299	&	2.30 	$\pm$	0.05 	&	164	$\pm$	4	&	2.71 	$\pm$	0.19 	&	42.37/38	\\
			11514	&	55301	&	2.04 	$\pm$	0.06 	&	118	$\pm$	4	&	2.82 	$\pm$	0.21 	&	32.45/28	\\
			11515	&	55303	&	2.19 	$\pm$	0.05 	&	136	$\pm$	4	&	2.63 	$\pm$	0.20 	&	55.49/55	\\
			11516	&	55306	&	2.06 	$\pm$	0.05 	&	115	$\pm$	4	&	2.70 	$\pm$	0.19 	&	46.57/55	\\
			11517	&	55321	&	2.25 	$\pm$	0.05 	&	161	$\pm$	4	&	2.83 	$\pm$	0.17 	&	32.93/38	\\
			11518	&	55325	&	2.26 	$\pm$	0.06 	&	117	$\pm$	4	&	2.04 	$\pm$	0.16 	&	108.11/92	\\
			11519	&	55327	&	2.23 	$\pm$	0.06 	&	109	$\pm$	3	&	2.00 	$\pm$	0.15 	&	79.73/89	\\
			11520	&	55330	&	2.19 	$\pm$	0.06 	&	101	$\pm$	3	&	1.94 	$\pm$	0.14 	&	81.00/86	\\
			13964	&	55899	&	2.16 	$\pm$	0.06 	&	120	$\pm$	4	&	2.41 	$\pm$	0.18 	&	23.70/26	\\
			13965	&	55982	&	2.14 	$\pm$	0.06 	&	107	$\pm$	4	&	2.23 	$\pm$	0.17 	&	79.64/86	\\
			14974	&	56273	&	2.20 	$\pm$	0.06 	&	100	$\pm$	4	&	1.93 	$\pm$	0.13 	&	82.19/76	\\
			14973	&	56363	&	2.19 	$\pm$	0.06 	&	105	$\pm$	4	&	2.03 	$\pm$	0.21 	&	62.64/79	\\
			16042	&	56652	&	2.15 	$\pm$	0.08 	&	71	$\pm$	3	&	1.45 	$\pm$	0.14 	&	51.79/55	\\
			16043	&	56749	&	2.09 	$\pm$	0.06 	&	113	$\pm$	4	&	2.53 	$\pm$	0.19 	&	74.20/88	\\
			17056	&	57008	&	2.25 	$\pm$	0.08 	&	103	$\pm$	4	&	1.83 	$\pm$	0.16 	&	19.63/18	\\
			17057	&	57100	&	1.97 	$\pm$	0.07 	&	105	$\pm$	5	&	2.90 	$\pm$	0.23 	&	17.31/16	\\
			18233	&	57441	&	2.26 	$\pm$	0.03 	&	59	$\pm$	1	&	1.03 	$\pm$	0.03 	&	129.62/124	\\
			18781	&	57442	&	2.22 	$\pm$	0.03 	&	60	$\pm$	3	&	1.12 	$\pm$	0.04 	&	84.04/77	\\
			18782	&	57444	&	2.23 	$\pm$	0.03 	&	62	$\pm$	1	&	1.13 	$\pm$	0.05 	&	55.06/66	\\
			18809	&	57459	&	2.24 	$\pm$	0.10 	&	58	$\pm$	4	&	1.05 	$\pm$	0.14 	&	34.69/39	\\
			18810	&	57460	&	2.26 	$\pm$	0.11 	&	61	$\pm$	4	&	1.07 	$\pm$	0.13 	&	30.13/39	\\
			18811	&	57461	&	2.31 	$\pm$	0.10 	&	61	$\pm$	4	&	0.99 	$\pm$	0.11 	&	35.67/39	\\
			18812	&	57463	&	2.26 	$\pm$	0.10 	&	59	$\pm$	4	&	1.04 	$\pm$	0.13 	&	39.33/41	\\
			18813	&	57464	&	2.18 	$\pm$	0.09 	&	61	$\pm$	4	&	1.18 	$\pm$	0.13 	&	30.50/41	\\
			18783	&	57498	&	2.33 	$\pm$	0.04 	&	52	$\pm$	1	&	0.83 	$\pm$	0.03 	&	117.91/108	\\
			18232	&	57505	&	2.18 	$\pm$	0.04 	&	63	$\pm$	2	&	1.24 	$\pm$	0.06 	&	78.34/76	\\
			18836	&	57506	&	2.18 	$\pm$	0.03 	&	64	$\pm$	1	&	1.25 	$\pm$	0.04 	&	133.45/134	\\
			18837	&	57508	&	2.38 	$\pm$	0.06 	&	55	$\pm$	2	&	0.80 	$\pm$	0.06 	&	47.90/47	\\
			18838	&	57536	&	2.35 	$\pm$	0.03 	&	51	$\pm$	1	&	0.79 	$\pm$	0.03 	&	160.11/137	\\
			18856	&	57551	&	2.36 	$\pm$	0.05 	&	49	$\pm$	1	&	0.74 	$\pm$	0.04 	&	149.33/124	\\
			19457	&	57799	&	2.23 	$\pm$	0.08 	&	83	$\pm$	4	&	1.37 	$\pm$	0.14 	&	58.57/52	\\
			19458	&	57800	&	2.22 	$\pm$	0.11 	&	66	$\pm$	4	&	1.23 	$\pm$	0.14 	&	41.37/43	\\
			20034	&	57854	&	1.97 	$\pm$	0.04 	&	115	$\pm$	3	&	3.10 	$\pm$	0.13 	&	95.16/102	\\
			20035	&	57857	&	2.10 	$\pm$	0.04 	&	103	$\pm$	3	&	2.26 	$\pm$	0.11 	&	91.83/84	\\
			20488	&	58122	&	1.97 	$\pm$	0.07 	&	131	$\pm$	7	&	3.59 	$\pm$	0.24 	&	40.20/41	\\
			20489	&	58198	&	2.01 	$\pm$	0.07 	&	113	$\pm$	2	&	3.01 	$\pm$	0.23 	&	72.90/69	\\
			21075	&	58230	&	1.98 	$\pm$	0.04 	&	175	$\pm$	5	&	4.59 	$\pm$	0.20 	&	85.46/101	\\
			21076	&	58232	&	2.00 	$\pm$	0.04 	&	190	$\pm$	5	&	4.93 	$\pm$	0.21 	&	118.29/103	\\
			21457	&	58569	&	2.13 	$\pm$	0.05 	&	83	$\pm$	3	&	1.75 	$\pm$	0.09 	&	144.15/123	\\
			21458	&	58570	&	2.16 	$\pm$	0.05 	&	88	$\pm$	3	&	1.76 	$\pm$	0.10 	&	119.93/117	\\
		\end{longtable}
	}

	\subsection{Clumpy Accretion Model Fitting}
	It can be clearly seen from the M87 black hole image published by the Event Horizon Telescope (EHT) in 2017 that the bright ring morphology appears to be inhomogeneous (\citealt{EHT+etal+2021}). Meanwhile, the year-scale variaton of M87 is a long-term evolution process, and the light variability characteristics are similar to the clumpy-ADAF model proposed by \citet{Wang+etal+2012}. 
	The accretion rates can be written in Eddington units, $\dot{M}=\dot{m} \dot{M}_{\rm Edd}$, where  $\dot{M}_{\rm Edd}$ is the Eddington accretion rates which can be defined as: 
	$\dot{M}_{\rm Edd}=L_{\rm Edd}/\eta c^2=2.2 \times 10^{-8} m\ \rm M_\odot\ {yr^{-1}}$ (\citealt{Matteo+etal+2003}), 
	$m=M_{\rm BH}/M_\odot$ is the dimensionless  mass of black hole,
	$c$ is the light speed, $\eta$ is the X-ray radiation efficiency ($\eta=0.1$ for standard models), $L_{\rm Edd}$ is the Eddington luminosity. 
	$\dot{m}$ can be expressed as $L/L_{\rm Edd}$.
	$L_{\rm Edd}=1.25 \times 10^{38} m\ {\rm erg\ s^{-1}}$ when  $\eta=0.1$ (\citealt{Wang+etal+2012}). 
	In M87 hot accretion flow, $\eta$ is about $10^{-5}$ (\citealt{Matteo+etal+2003}).
	The luminosity of black hole is $L=4\pi r^2\ F\ {\rm erg\ s^{-1}}$, where $F$ is flux, $r$ is the distance of M87. 
	Then we can get the mass accretion rate of M87 as:
	\begin{equation}\label{eq2}
		\dot{M}=\frac{L}{L_{\rm Edd}} \times \frac{\eta_{10^{-5}}}{\eta_{0.1}} \times \dot{M}_{\rm Edd},
	\end{equation}
	where $\eta_{0.1} = \eta/0.1$ and $\eta_{10^{-5}} = \eta/10^{-5}$. With this formula, we find that the X-ray luminosity is in proportion to the mass accretion rate which is also mentioned by \cite{Ishibashi+Courvoisier+2009}.
	
	In the past, both theory and observation supported that the accretion flow around black hole in LLAGNs was inhomogeneous (\citealt{Celotti+Rees+1999}). Due to thermal instability and viscous instability, it would create cold clumps in the disc (\citealt{Krolik+1998}) and then fallback into the central black hole. During the process of accretion, the clump will be disrupted by tidal force and release a burst of energy (\citealt{Celotti+Rees+1999}; {\citealt{Strubbe+Quataert+2009}}). As for the mass accretion rate of clumpy gas, {\citet{Xiang+Cheng+2020}} derived a solution as follows:
	\begin{equation}\label{eq3}
		\dot{M}(x,\tau)\ =\ \dot{M}_0\ (\frac{1}{2x}\ -\ \frac{2x}{\tau})\ \frac{x^{3/4}}{\tau} e^{-\frac{1+x^2}{\tau}}I_{1/4}\ (\frac{2x}{\tau}),
	\end{equation}
	and
	\begin{equation}\label{eq4}
		\tau\ =\ (t-t_0)/\tau_0,
	\end{equation}
	where $\dot{M}_0 = 6\pi\nu\Sigma_0$, $\Sigma_0$ is the initial surface density of the clumpy gas \citep{Xiang+Cheng+2020}, $\nu$ is kinematic viscosity parameter \citep{Lin+1987},
	$x$ can be written in $R/R_0$ and represents the dimensionless distance to central BH and $R_0$ is the radius where the clump forms, $R_0$ is about $100R_{\rm Sch} \sim 1000R_{\rm Sch}$ \citep{Wang+etal+2012}, $t_0$ stands for the start date of the clumpy accretion, $\tau_0$ is the time-scale of gas falling and $I_{1/4}$ is the modified Bessel function.
	
	From 2007 to 2008 data, \citet{Xiang+Cheng+2020} found there was an obvious anti-correlation between photon index and flux. Based on this characteristic, \cite{Xiang+Cheng+2020} divided the states of nucleus into two types and those five points with lower flux and higher index were defined as the ‘first class’; three points with higher flux and lower index were defined as the ‘second class’. The first class of low state was corresponding to the pure ADAF model (\citealt{Li+etal+2009}) and had been successfully fitted by clumpy accretion model (\citealt{Xiang+Cheng+2020}). The second class of high state could be separated into two components, an ADAF component and a flaring component, and the former might also match the value of disc evolution. 
	However, this classification method can not work well for the  long-term light curve. It can be seen from the right panel of Fig.~\ref{Fig2}  that there is no obvious bimodal structure in the histogram distribution of photon index and flux in the 12 years observations. 
	In order to get the year-scale variation in the light curve, we propose a  universal classification method for this two states. The specific classification steps are as follows:
	
	Select all points within a half year period around each point. If there are no points in the range, select the two nearest points on the left and right. We check whether this point is the maximum value in this range and is 50\% higher than the average value of the rest points. For points satisfying the condition, it is considered to be at flaring state. After removing these points, this calculation process is repeated until there are no new flaring state points identified. 
	
\begin{figure}
		\centering
		\includegraphics[width=13cm, angle=0]{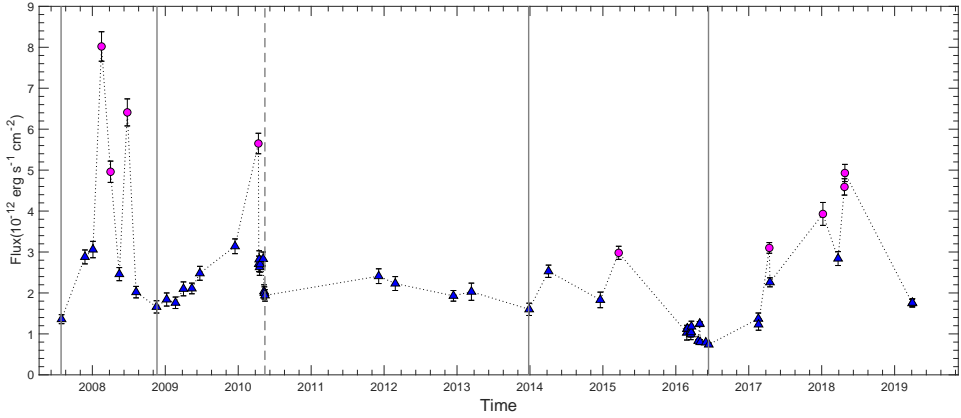}
		\caption{The light curve of the core. {\it Blue triangles}: the observations of non-flaring state; {\it magenta dots}: the observations of flaring state. The gray lines represent the positon of the local minimum brightness.}
		\label{Fig3}
	\end{figure}

	The results obtained by this method are shown in Fig.~\ref{Fig3}. 
	These  blue triangle points are classified into non-flaring state which are considered to be accompanied by the clumpy accretion activities. 
	These points marked in magenta dots are at flaring state which are related to the flare events. Meanwhile, the distinguished flaring state and non-flaring state from 2007 to 2008 are consistent with the classification results in \citet{Xiang+Cheng+2020}.

\begin{figure}
	\centering
	\includegraphics[width=15cm, angle=0]{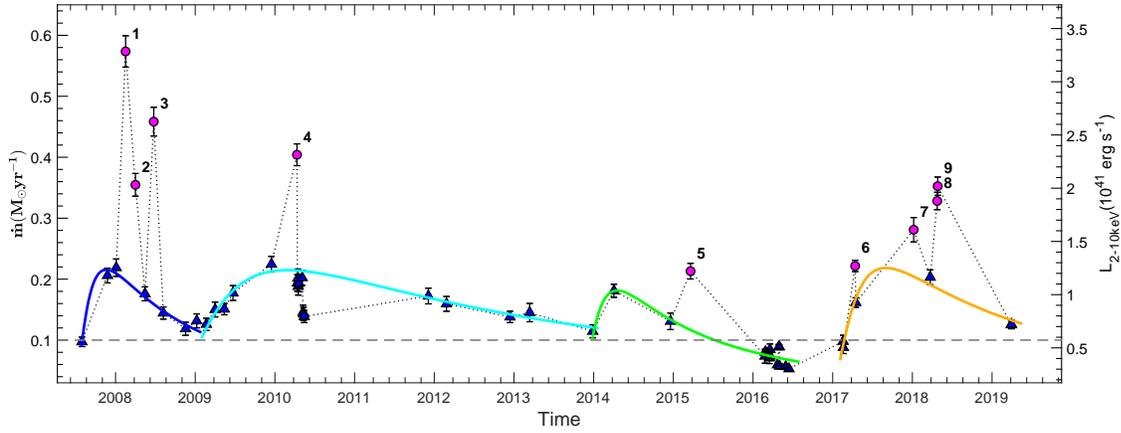}
	\caption{The long-term evolution of the mass accretion rate for the central BH of M87. {\it Blue filled triangles}: non-flaring state; {\it  magenta filled dots}: flaring state. {\it The solid lines} represent the four clumpy accretion model fitting results. {\it The gray dotted line} represents the mass accretion rate of $ 0.1 \rm M_{\odot} yr^{-1}$. } \label{Fig4}
\end{figure}

To meet the conditions of the model fitting, it is necessary for us to divide the starting and ending time of the accretion process from the long-term observations.
After excluding those flaring state points, we find that there are five local minimum of brightness (segmented by the gray line in Fig.~\ref{Fig3}).
For the clumpy accretion model, the light curve will experience a rapid increase and then slow decrease during the accretion process (\citealt{Wang+etal+2012}). However,  the decline rate after the flare event in April 2010 is much faster than the ascent rate of 2009, which is inconsistent with the physical characteristics described by the accretion model.
Therefore, we do not take it as the beginning or end of the accretion process (the position is shown by gray dotted line in Fig.~\ref{Fig3}).
Based on this standard, 4 candidate clumpy accretion components are identified. 
The first component contains the entire time period of the accretion process in \citet{Xiang+Cheng+2020}.
Fitted these candidate accretion parts with formula (3), we get the results shown in Fig.~\ref{Fig4} and the fitting parameters are listed in Table~\ref{Tab2}. 
Meanwhile, we label the nine observations of flaring state with a sequence number.

\begin{table}
	\begin{center}
		\caption[]{ The parameters of the clumpy accretion fitting results.  $M_{\rm c}$ and $R_{\rm c}$ is the mass of gas clump and radius of clump, respectively. The mass of clump is the result of the integration of $\dot{M}$ by time.  }\label{Tab2}
		
		\setlength{\tabcolsep}{8pt}
		
		\begin{tabular}{lllllll}
			\hline\noalign{\smallskip}
			& $\dot{M}_0$ & $x$ &  $\tau_0$ & $t_0$  & $M_{\rm c}$  & $R_{\rm c}$ \\
			& ($\rm M_\odot \rm yr^{-1}$) &   &(days) &  (date)   & ($\rm M_\odot$) &($10^{13} \rm cm$) \\
			\hline\noalign{\smallskip}
			fit1      &  1.04       & 0.04     &      225      &2007-05-23  &  0.23 &   8.62        \\ 		
			fit2      &  1.03        & 0.02     &     819    &2008-05-12    &  0.83  &   13.33       \\ 		
			fit3     &  0.87       & 0.02    &      253     &2013-09-27     & 0.30   &   9.45        \\ 		
			fit4      &  1.05       & 0.03  &  387      &  2016-10-23       &  0.38  &   10.23   \\ 		
			\noalign{\smallskip}\hline
		\end{tabular}
	\end{center}
\end{table}

\section{Discussion}
\label{sect:discussion}

\subsection{The Physical Characteristics of Clumpy Accretion}
In Fig.~\ref{Fig4}, it shows that those non-flaring state observations  are basically on the fitting curve which illustrate that our classification method for clumpy accretion is reasonable. Although the luminosity dropped sharply to a very low state after the flare in 2010, it didn't bring a great influence to the long-term evolution of the disc. We give a more detailed explanation of this phenomenon in Section 4.3.

\begin{figure}
	\centering
	\includegraphics[width=10cm, angle=0]{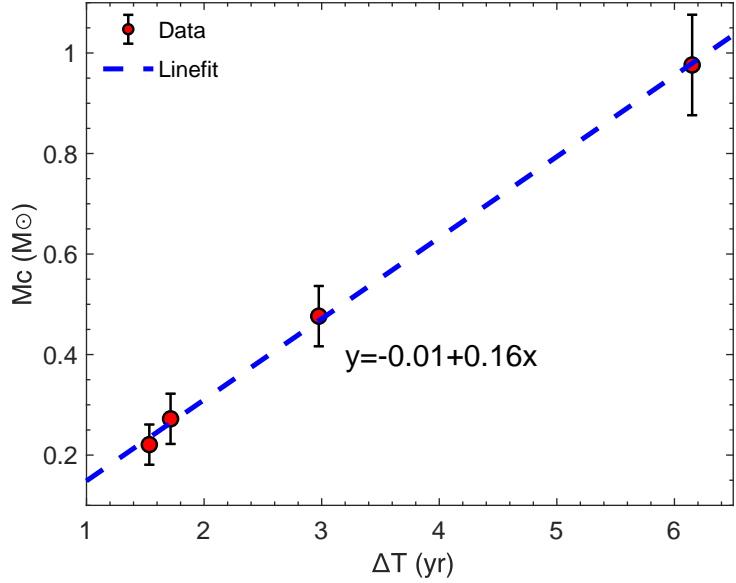}
	\caption{The variation of clump mass with accretion time-scale. The red dot represents the mass of each clump; the dashed blue line is the linear fitting result. The error of the clump mass is calculated based on the 95\% confidence limits of the model fitting result in Section 3.2.}
	\label{Fig5}
\end{figure}

Based on the parameters listed in Table~\ref{Tab2}, we can find that all of the dimensionless distance to central BH $(x)$ are no more than 0.05, which indicates that the region where the clumps form is very close to the black hole. 
Our fitting results for the first clump are consistent with the results in \citet{Xiang+Cheng+2020}.
Although the size of the source region adopted in this paper ($0.8^{\prime \prime} \times 2.6^{\prime \prime}$) is smaller than that in \citet{Xiang+Cheng+2020} ($1.8^{\prime \prime} \times 2.3^{\prime \prime}$) which leads to higher luminosity and mass accretion rate, it doesn't change the position where the clump forms.
Meanwhile, it can be seen from Fig.~\ref{Fig4} that the accretion  on the black hole is discontinuous and  the size of the clump is randomly generated. Since the solution of the clumpy accretion is a function of mass accretion rate with time, the mass of the clump could be obtained through integration of $\dot{M}$ by time. 
When the mass accretion rate is lower than $ 0.1 \rm M_{\odot}  yr^{-1}$,  the radiation generated by accretion is very weak.
Therefore, we take $ 0.1 \rm M_{\odot}  yr^{-1}$ as the minimum threshold of the clumpy accretion.  We define the time range above this threshold as the accretion time-scale ($\Delta T$) of a clumpy accretion.
Based on this standard, we find that the accretion of the last gas clump had not been completed within the selected observation time range. According to the results predicted by the model, the accretion rate would drop to $ 0.1 \rm M_{\odot}  yr^{-1}$ on February 11th, 2020.
Then we get the mass of each clump by integral calculation.
As the morphology of the clump is spherical, the radius of a clump could be estimated by 
$R_{\rm c}$ = (3$M_{\rm c}$/(4$\pi n_{\rm cl}$$m_{\rm p}$))$^{\frac{1}{3}}$, where $m_{\rm p}$ is the mass of proton, $n_{\rm cl}$ is the density of gas clump and the typical density is $\sim 10^{14} \rm \ cm^{-3}$ (\citealt{Xiang+Cheng+2020}). 
The values of $M_{\rm c}$ and $R_{\rm c}$ are listed in Table~\ref{Tab2}, and the linear relationship between $M_{\rm c}$ and  $\Delta T$ is given in Fig.~\ref{Fig5} (the dashed blue line). The regression equation is $\rm M_c$ $= q_0 + q_1 \Delta \rm T$ and their correlation coefficient is 0.98. The fitting parameter: $q_1 = 0.16 \pm 0.01$.
From the linear fitting result, we can deduce that the time-scale of clumpy accretion is determined by the size of the gas clump. With a mass of $\sim 0.16 \rm M_{\odot}$, the accretion process will last for about one year and lead to the variation of the X-ray luminosity.

\subsection{The $\bf \Gamma$ — $\bf  F_{2-10KeV}$ Correlation Between Flaring State and Non-flaring State}

\begin{figure}
	\centering
	\includegraphics[width=10cm, angle=0]{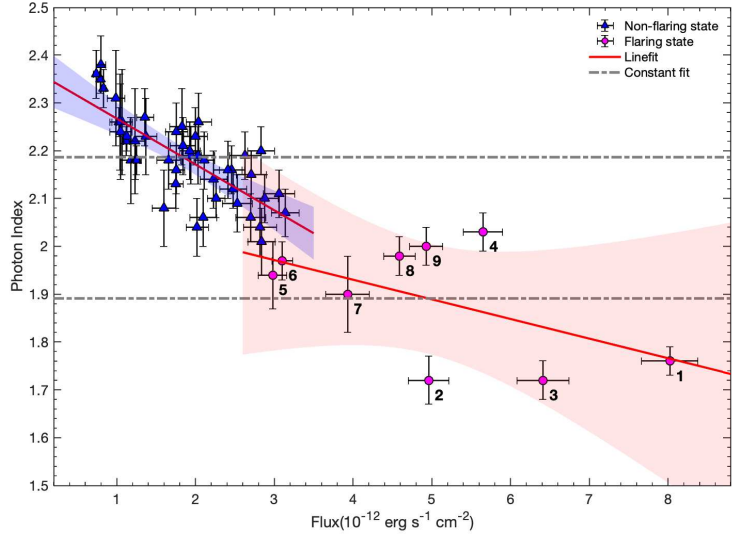}
	\caption{Distribution of photon index versus flux. Blue triangles are the non-flaring state observations;  magenta dots are the flaring state observations. The sequence numbers are corresponding to those in Fig.~\ref{Fig5}. The dashed grey line represents a constant fit, and the red solid line represents a linear fit of the form 
		$\Gamma = p_0 + p_1 \Phi_0$, where $\Phi_0$ is the spectral flux; the filled region represents the linear fit in $95 \%$ confidence limits. The value of $p_1$ is given in the text.}
	\label{Fig6}
\end{figure}

The correlation between the photon index and flux in very-high-energy (VHE) gamma-rays of M87 have been discussed by \citet{Acciari+etal+2008}.  Due to the small number of observations, no obvious correlation was found.
However, the long-term X-ray observations provides enough data for us to study the relationship between this two parameters.
In order to test if there is any difference of the $\Gamma$ — $\rm  F_{2-10keV}$ distribution between the flaring state and non-flaring state, we take the photon index ($\Gamma$) and the flux of the two states respectively. 
Then fitting these two parameters with a linear test function $\Gamma = p_0 + p_1 \Phi_0$. 
The results are shown in Fig.~\ref{Fig6}. The $\chi^2 / dof$ of the linear fit of the non-flaring state is 43.68/45, with the parameter $p_1 = -0.10 \pm 0.01$; the $\chi^2 / dof$ of the linear fit of the flaring state is 6.39/7, with the parameter $p_1 = -0.04 \pm 0.02$.  However, as $p_1$  is consistent with zero, a constant photon index fit is applied to compare with the linear fitting results. The constant fit gave a $\chi^2 / dof$ of 97.75/46 and 8.87/8 for the non-flaring state and flaring state, respectively. 
As a consequence, the fitting result shows that the flux is inversely proportional to the photon index for the non-flaring state (clumpy accretion components).
For the flaring state (flare components), due to the $\chi^2 / dof$ of the linear fit is consistent with the $\chi^2 / dof$ of constant fit, no significant evidence is provided for there is any correlation between the two parameters. 

The anti-correlation between photon index and flux in  X-ray band is also predicted by ADAF model (\citealt{Yuan+etal+2007}). For LLAGNs, the X-ray emission is mainly dominated by the Comptonization of the hot gas in ADAF (\citealt{Gu+Cao+2009}; \citealt{Xiang+Cheng+2020}).
However, according to the distribution of the two parameters, we can see that there is a cross connection between the non-flaring state and flaring state.
For the flare in 2008, the peak flux reached 8.02$\times$10$^{-12}$ erg s$^{-1}$ cm$^{-2}$,  with the photon index of 1.76 (number 1 in Fig.~\ref{Fig4} and Fig.~\ref{Fig6}). Then the intensity dropped to 4.96$\times$10$^{-12}$ erg s$^{-1}$ cm$^{-2}$, and the photon index was 1.72 (number 2 in Fig.~\ref{Fig4} and Fig.~\ref{Fig6}), almost the same as before.
This shows that the event dominated by the flare varies greatly of the flux intensity, but keep the same feature of the photon index as the flare.
Meanwhile, $\Gamma$ -- $\rm F_{2-10keV}$ distribution of the flaring states  seems to present two branches. Such branches may imply different origin mechanism of the flares. 
Carry out high-frequency follow-up observation after the flare events may help us to understand the physical mechanism of this process.

\subsection{Why Can the Luminosity be Lower Than the Clumpy Accretion Component? }
It can be seen from the distribution of light curve in Fig.~\ref{Fig4} that the flare often occurs at a high mass accretion rate. After the intensity reaches the peak, a sudden decrease in luminosity will be accompanied within a few days. The flare event in 2010 was particularly representative. The luminosity declined by 52.04\% in two days after the peak intensity (number 4 in Fig.~\ref{Fig4} ).
At this time, the luminosity was consistent with that of the non-flaring state. And then the luminosity decreased by 28.41\% over the next 30 days.
A light variability in 2017 is also very similar to this situation. We find that after the flare (number 6 in Fig.~\ref{Fig4} ), the luminosity decreased by 27.10\% within three days which was consistent with the ADAF components. Due to the lack of observations afterwards, it can not be sure whether the luminosity will continue to decrease for a period. But it is not a coincidence that the luminosity  changes rapidly in a short time after the flare. Then, we analyze the origin mechanism of the flare and put forward a possible explanation for the above phenomenon.

The M87 image captured by EHT presents  that the inhomogeneous ring-like structure  seems to be clumpy. 
Meanwhile, the polarization map of M87 shows that there is strong magnetic field around the black hole 
(\citealt{Polarization+etal+2021}), which is closely related to the accretion mode of the black hole.  Now it has been confirmed that the accretion flow in M87 is magnetically-arrested disk (MAD, \citealt{Xie+etal+2019}; \citealt{Magenetic+etal+2021}), and it is based  on general relativistic magnetohydrodynamic simulations (GRMHD, \citealt{McKinney+etal+2012}; \citealt{Yuan+Narayan+2014}). In MHD model, there are magnetic arcades emerging from the disc into corona (\citealt{Yuan+etal+2009}). The formed flux ropes keep a balance between the magnetic compression and magnetic tension. Nevertheless,the equilibrium will be broken down by the turbulence in the photosphere and leads to rapid magnetic reconnection (\citealt{Lin+etal+2003}; \citealt{Yuan+etal+2009}).In this process, part of the energy transfers into the kinetic power of the plasma to ignite the flares, and part of it  pushes the mass through the corona. 

Our analysis shows that the flare might be triggered by magnetic reconnection, and the huge energy released from the process could blow material away from the accretion disc. As the structure of the accretion disc is destroyed, the luminosity decreases rapidly.
However, with the accretion of the gas clump, the damaged part will be refilled. As a consequence, the luminosity returns to normal. This further shows that the flare does not have a great impact on the overall evolution of the accretion disc, which is consistent with the conclusion in Section 4.1.

\section{Conclusions}
\label{sect:conclusion}
We search the long-term X-ray variation of M87 from {\it Chandra} archival data. In our analysis, 56 observations from 2007 to 2019 are adopted. We distinguished the `non-flaring state' from `flaring state' with a universal classification method. The evolution of the non-flaring states could be well explained by the accretion of gas clumps. We also discussed the physical characteristics of clumpy accretion. The main results are listed as follows:

(i) From 2007 to 2019, the central black hole of M87 have accreted 4 gas clumps. The time-scale of accretion is determined by the size of the clump. Generally, it takes about one year to complete the accretion process of a mass of $\sim \rm 0.16 M_{\odot}$.

(ii) We analyze the correlation of photon index against flux between the non-flaring state and flaring state. By linear fitting, we find that there is a significant anti-correlation between the two parameters of non-flaring states. However, the correlation is not significant for flaring states. 

(iii) The flare always occurs at a high mass accretion rate. 
Aftering the flare, there could be a steep luminosity drop to a level lower than that of the ADAF components. This hints that there might be a strong magnetic field around the black hole and flares could be related to the magnetic reconnections. The energy released by this process might temporarily destroy the structure of the disc. However, with the accretion of gas clumps, the damaged part could be filled again, and then the system returns to normal.

\begin{acknowledgements}
	We thank the anonymous referee for detailed and constructive suggestions. 
	This work is supported by the National Science Foundation of China (grant 11863006, U1838203, U2038104), the Science \& Technology Department of Yunnan Province - Yunnan University Joint Funding (2019FY003005), and the Bureau of International Cooperation, Chinese Academy of  Sciences under the grant GJHZ1864. We thank Paolo Tozzi for his helpful comments.
	
\end{acknowledgements}

\bibliographystyle{raa}
\bibliography{bibtex}

\label{lastpage}

\end{document}